\documentclass[conference]{IEEEtran}
\IEEEoverridecommandlockouts
\usepackage{cite}
\usepackage{amsmath,amssymb,amsfonts}
\usepackage{algorithmic}
\usepackage[]{algorithm2e}
\usepackage{graphicx}
\usepackage{textcomp}
\usepackage{xcolor}
\def\BibTeX{{\rm B\kern-.05em{\sc i\kern-.025em b}\kern-.08em
    T\kern-.1667em\lower.7ex\hbox{E}\kern-.125emX}}
    
\usepackage{filecontents}

\usepackage{grffile}

\usepackage{ulem}
\SetKwComment{Comment}{/* }{ */}

\newcommand{\todo}[1]{\textbf{TODO: #1}}
\usepackage{tikz}

\usepackage{pgfplots}
\pgfplotsset{compat=newest}
\pgfplotsset{
	scriptsize/.style={
		width=4.5cm,
		height=,
		legend style={font=\scriptsize},
		tick label style={font=\scriptsize},
		label style={font=\footnotesize},
		title style={font=\footnotesize},
		every axis title shift=0pt,
		max space between ticks=15,
		every mark/.append style={mark size=7},
		major tick length=0.1cm,
		minor tick length=0.066cm,
	},
}
\pgfplotsset{legend cell align=left}
\pgfplotsset{xmajorgrids}
\pgfplotsset{ymajorgrids}
\pgfplotsset{scale only axis}
\definecolor{matlab1}{rgb}{0,0.4470,0.7410}
\definecolor{matlab2}{rgb}{0.8500,0.3250,0.0980}
\definecolor{matlab3}{rgb}{0.9290,0.6940,0.1250}
\definecolor{matlab4}{rgb}{0.4940,0.1840,0.5560}
\definecolor{matlab5}{rgb}{0.4660,0.6740,0.1880}
\definecolor{matlab6}{rgb}{0.3010,0.7450,0.9330}
\definecolor{matlab7}{rgb}{0.6350,0.0780,0.1840}
\pgfplotscreateplotcyclelist{matlab}{
	{matlab1,solid},
	{matlab2,dashed},
	{matlab3,dashdotted},
	{matlab4,dotted},
	{matlab5,densely dashed},
	{matlab6,densely dashdotted},
	{matlab7,densely dotted}
}
\pgfplotsset{cycle list name=matlab}
\pgfplotsset{every axis plot/.append style={line width=1pt}}
\pgfplotsset{/pgf/number format/.cd,1000 sep={\,}}

\begin{document}

\title{Leveraging the SCION Internet Architecture to Accelerate File Transfers over BitTorrent}

\author{\IEEEauthorblockN{Marten Gartner}
\IEEEauthorblockA{\textit{Otto-von-Guericke University} \\
Magdeburg, Germany \\
marten.gartner@ovgu.de}
\and
\IEEEauthorblockN{Thorben Krüger}
\IEEEauthorblockA{\textit{Otto-von-Guericke University} \\
Magdeburg, Germany \\
thorben.krueger@ovgu.de}
\and
\IEEEauthorblockN{David Hausheer}
\IEEEauthorblockA{\textit{Otto-von-Guericke University} \\
Magdeburg, Germany \\
hausheer@ovgu.de}
}
\maketitle
\begin{abstract}

As the needs of Internet users and applications significantly changed over the last decade, inter-domain routing became more important to fulfill these needs. The ways how data flows over the Internet are still completely in the hand of network operators, who optimize traffic according to their own, local view of the network. We observe two potential limitations from this: Optimizing according to the local view may a) result in unused capacities in the global network and b) not meet the actual needs of users and applications. To identify and overcome these limitations, we present our BitTorrent over SCION approach, which enables multipath communication and intelligent path selection for endhosts in global torrent networks. We compare our implementation against BitTorrent over BGP and BGP-M in a small-scale Internet topology, observing an increase in goodput of 48\% through multipathing compared to BitTorrent over BGP and 33\% compared to the BGP-M candidate. Furthermore, we show that our proposed disjoint path selection algorithm is able to improve traffic flow in the network with a low number of outgoing connections to unchoked peers.

\end{abstract}


\section{Introduction}
The Internet was designed decades ago, and some of the original design decisions have had unfortunate side-effects and security implications that persist to this day. So far, from the perspective of a mere host, there has been no general mechanism for splitting traffic across multiple different paths to a specific destination, which would be beneficial bandwidth-intensive applications.

As the de facto standard protocol for inter-domain routing, BGP \cite{rekhter1995rfc1771} is used to disseminate routes between Autonomous Systems (ASes), forming the Internet as we know it today. Within each AS on a given path through the Internet, packet forwarding is affected by (sometimes highly complex) local traffic engineering preferences and policies. In general however, there is just a single packet forwarding path between two hosts in a BGP-based Internet. To also add support for multiple forwarding paths, BGP-M \cite{li2022bgp} was proposed, which adds support for load sharing across multiple inter-domain links, subject to configurable preferences. However, BGP-M is does little to help optimizing traffic flow in the global Internet: Firstly, it's impact is necessarily of limited, local scope. Secondly, BGP-M is not adaptive and can not dynamically react to changing network conditions. Given these limitations and given the existence of alternative inter-domain paths in the Internet \cite{bakhshaliyev2019investigating}, we hypothesize that the current Internet still has unused capacities that cannot be exploited entirely by using BGP or BGP-M as the inter-domain routing protocol. 

Path-aware networking architectures promise to overcome these limitations by offering deeper insight into and better control over packet forwarding in the network. Different approaches to enable path-control in networks have been proposed. Some try to work within the limits of the current Internet architecture \cite{xu2012mpath, gvozdiev2018low}, other attempts involve a complete redesign of the Internet architecture from scratch. While there are a number of path-aware approaches \cite{yang2003nira, rothenberger2020piskes} with their own merits, our work focuses on the (arguably) most mature and most widely-deployed open-source SCION architecture \cite{chuat2022complete, zhang2011scion}.


To leverage unused capacities in the network, especially in the backbone, Peer-to-Peer \emph{(P2P)} applications promise unique opportunities through their globally distributed nature \cite{rumin2011deep, cuevas2013bittorrent, wang2009locality}. As the most well-researched and understood protocol, we selected BitTorrent as the foundation of our work, adding support for active endhost-based path-selection via SCION. We compare our augmented BitTorrent implementation with a non path-aware BitTorrent implementation in comparable BGP and BGP-M-based inter-domain network topologies. 

We anticipate that any achieved improvements will easily translate to other P2P networks, e.g. IPFS \cite{trautwein2022design}. While BitTorrent itself is famous for improving bandwidth utilization on the last mile, we hypothesize that in combination with SCION, it could also unlock unused capacities in the network at large. To this end, we contribute the following:
\begin{itemize}
    \item We discuss the existence of unused capacities in the backbone with BGP or BGP-M as utilized inter-domain routing protocol
    \item We show why path-aware networking is able to unlock these capacities and discuss the impact of host-based path selection
    \item We present our BitTorrent over SCION design introducing the notion of path-level peers and propose an algorithm for disjoint path selection
    \item Finally, we analyze the performance improvements of BitTorrent over SCION aggregating unused capacities in the network compared to BitTorrent over BGP and \\ BGP-M
\end{itemize}

The remainder of this work is structured as follows: In Section \ref{sec:background} we provide background for SCION and BitTorrent, followed by a discussion about limitations of BGP-based deployments and impacts of host-based path selection in Section \ref{sec:bgp-limits-and-scion-poweders}. We present the design and implementation of multipath support for BitTorrent over SCION in Section \ref{sec:multipathbt}, followed by a presentation of our disjoint path selection algorithm. Afterwards, we show our virtualized Internet-scale testbed in Section \ref{sec:benchmark-setup} and the experimental results of comparing Multipath BitTorrent over SCION against an unmodified implementation in BGP and BGP-M-based scenarios in Section \ref{sec:eval-large}. Finally, we discuss related work in Section \ref{sec:relwork}, conclude and provide outlook for future work in Section \ref{sec:conclusion}.


\section{Background}
\label{sec:background}
In the following, we provide a brief overview on BGP, BGP-M and the SCION architecture as well as on BitTorrent.

\subsection{BGP and BGP-M}
BGP plays a unique role in the current Internet. While it can also enable in intra-domain routing (iBGP), we will exclusively refer to BGP's more important role as the Internet's predominant inter-domain routing protocol in the rest of this work. The principle task of a BGP border router is to inform and update its neighbours about specific IP address ranges (i.e., IP prefixes), to which the router's AS is able to forward traffic to. The router announces the prefixes that its AS has learned from its other neighbors. Routing loops are prevented by means of the \emph{AS\_PATH}, a list of hops to which each router prepends its own AS Number (ASN) before sending it on as part of a route announcement. BGP routers maintain the respective AS\_PATHs as well as the announced prefixes in a routing table. The forwarding destination is determined by referencing this table based on the destination address of an incomping packet. By default, BGP selects a single path for each match in the routing table according to configurable policies. These policies may reflect e.g., local forwarding preferences, filters constraining the number of exported routes, etc. 

To account for the fact that there often are multiple matching paths to a particular destination, multipath BGP \emph{(BGP-M)} \cite{fujinoki2008multi} was introduced. For certain cases, BGP-M adds support for traffic load splitting\footnote{Load splitting is based on flow hashes of the 5 tuple \emph{(src address, dst address, src port, dst port, layer 4 protocol)}.} over multiple outgoing AS links. In the most simple case, if several possible forwarding AS\_PATHs are of the same length, Equal-Cost Multipath Routing \emph{(ECMP)} can be applied, with a configurable maximum number of parallel paths\footnote{In some border router implementations, the relevant settings are: a) \texttt{bgp bestpath as-path multipath-relax} to enable load splitting for equal length paths, and b) \texttt{maximum-paths} to set the maximum number of paths over which load splitting will be performed.}. In Addition, BGP-M also allows for more advanced policies on how to combine multiple paths, e.g., some Cisco routers offer Unequal-Cost Load Sharing \cite{ucmpbgp}, which allows load splitting over paths with different length by using a dedicated configurable weight.

\subsection{SCION}

The SCION architecture \cite{zhang2011scion} has been designed to overcome the limitations of BGP and BGP-M in the future Internet, addressing modern threat models at the fundamental protocol level and endeavouring to avoid many current issues with hijacking attacks and single roots of trust. SCION also promises communication guarantees and path control capabilities, allowing applications to use two or more paths in parallel to a given destination, generally enabling \emph{multipath communication}.

For traditional multipath approaches like MPTCP \cite{peng2014multipath} and MPQUIC \cite{de2017multipath} to work as intended, a host must provide multiple network interfaces. With these protocols, multipath communication implies sending data over multiple interfaces in parallel, beyond which no further influence on the data forwarding paths is possible. SCION on the other hand is based around "packet carried forwarding state" (PCFS), where every packet contains the complete inter-AS path to the intended destination (in the form of \emph{hop fields}) in its packet header. Packets can thereby be easily directed via different paths, by changing the SCION header alone. Our work heavily relies upon this precise \emph{path awareness} property of the SCION architecture.

Unlike the flatter organizational hierarchy of today's BGP-based Internet, in SCION, collections of ASes are combined into Isolation Domains, \emph{(ISDs)}, which are envisioned to correspond to, e.g., geographical regions or legislative domains (like a single country), but can also be made up of company or research networks. One or multiple ASes in an ISD form the \emph{ISD Core}, which collectively manages a cryptographic trust root (\emph{TRC}) on behalf of the other ISD members, enabling service authentication and other cryptographic functions within the ISD, simultaneously avoiding many of the notorious problems that plague globally centralized cryptographic trust systems with single points of failure that are beyond the control of the ISD. The ISD Core also manages the exchange of path information with other ISDs, while independent peering is also possible among non-Core ASes of different ISDs. 

\subsection{BitTorrent}

The BitTorrent protocol specifies file transfer as a distributed mechanism between peers without the need for any central coordination. Some initial way to exchange network addresses among peers is nevertheless required, e.g., by  means of a \emph{tracker} or, alternatively, a distributed hash table (DHT). In BitTorrent, files are typically not transferred in the usual form of a single, continuous byte stream that contains the complete file. Instead, large files are split into equal-sized \emph{pieces} (typically with a size between 32KB and 256KB). It is a key feature of BitTorrent that exchange of these pieces can be easily parallelized. BitTorrent peers that already have a complete local copy of a file are referred to as \emph{seeders}, as opposed to \emph{leechers}, which still have to obtain some or all of the constituent pieces of the file from other peers. For each new file uploaded to the BitTorrent network, there must be at least one seeding peer that initiates the distribution of the file. 



\section{Host-Based Path Selection}
\label{sec:bgp-limits-and-scion-poweders}

\subsection{Limitations of the BGP-Based Inter-Domain}
Generally, two critical aspects of BGP and BGP-M for inter-domain routing may lead to unused capacities in the global Internet: Firstly, AS operators inherently only have limited insight into network conditions beyond their local network and can also only manage traffic locally. Secondly, BGP and BGP-M do not provide features to dynamically adapt routing to different needs. In this section, we further characterize these limitations before discussing possible improvements that path-aware networking could bring to the table.

Despite the large benefits of traffic engineering of AS-operators on intra-domain level, the potential of optimizing inter-domain routing is limited in the current Internet. Each operator can only optimize the traffic flow in their own AS until it reaches its local destination or the neighbour AS. This may especially impact performance for flows that traverse multiple hops before they reach their destination, since each hop performs its own, local optimization. In case one of the first hops performs a non-optimal routing decision for the flow (e.g. routing to particular neighbour interfaces that are already under heavy load), the overall performance of the flow is affected.

As discussed in Section \ref{sec:background}, BGP-M provides several options to perform load sharing on multiple links. However, these options need to be configured statically in the network. Consequently, the network can not always fulfill the varying needs of different participants, e.g., endhosts who prefer to optimize for different criteria. BGP-M reflects the anticipated needs of AS operators, not the actual needs that endhosts have, which may differ significantly from those anticipated by the AS operators.

Path-aware networking promises to overcome many of these limitations and their impacts on performance, and, (in the case of SCION,) gives endhosts the opportunity to freely choose suitable inter-domain paths for their traffic. 


\subsection{Implications of Host-Based Path Selection}

Traffic engineering on the Internet is performed by network operators attempting to locally optimize data flows for various factors. With host-based path selection, operators hand over this control to endhosts. While this promises to help endhosts to optimize their traffic, it comes with potential implications for network operators and may be at odds with their own interests, especially with respect to the inter-domain. Operators tend to prefer the use of peering links over that of transit links to avoid costs, while endhosts do not have such an incentive to avoid transit links. Additionally, host-based path selection ideally requires up-to-date insight into network conditions on all hosts. It also needs to ensure that hosts do not change paths too often, which could result in undesirable oscillation. In simulation, Scherrer et al. show that the impact of oscillation through host-based path selection is low\cite{scherrer2022axiomatic}. Moreover, in SCION, endhosts can only dictate the ingress and egress router interfaces of ASes, allowing network operators to still optimize their internal traffic within these constraints. Overall, while the benefits, drawbacks and trade-offs of path-awareness are certainly not yet fully understood, there is significant potential for it in the the future Internet.



\section{Design and Implementation of BitTorrent over SCION}
\label{sec:multipathbt}

In this section, we present our approach of path-level peers to enable multipath support for BitTorrent over SCION, followed by our disjoint path selection algorithm.

\subsection{Multipath through Path-Level Peers}


As one of its features, SCION provides path control for inter-AS traffic while guaranteeing that the traffic flows along the chosen paths. This opens up the potential to aggregate capacities in the network, enabling applications to leverage multipath communication and parallel data processing to increase performance. In this work, we choose BitTorrent as a suitable, already parallelized application as a foundation for our experiments on bandwidth aggregation via multiple SCION paths.

By default, a BitTorrent peer is identified solely by its network address and port. This address could be either an IPv4, IPv6 or, in our case, a SCION address. We will refer to peers that are only described by their address as \emph{address-level} peers for the rest of this work. To distinguish between multiple SCION paths to a particular peer, we introduce a new representation that we will refer to as a \emph{path-level} peer. They are represented by the tuple $(addr, path)$, consisting of the peer's SCION address (including the port), together with one possible path to this address.

\begin{center}
	\begin{figure}
		\includegraphics[width=0.5\textwidth]{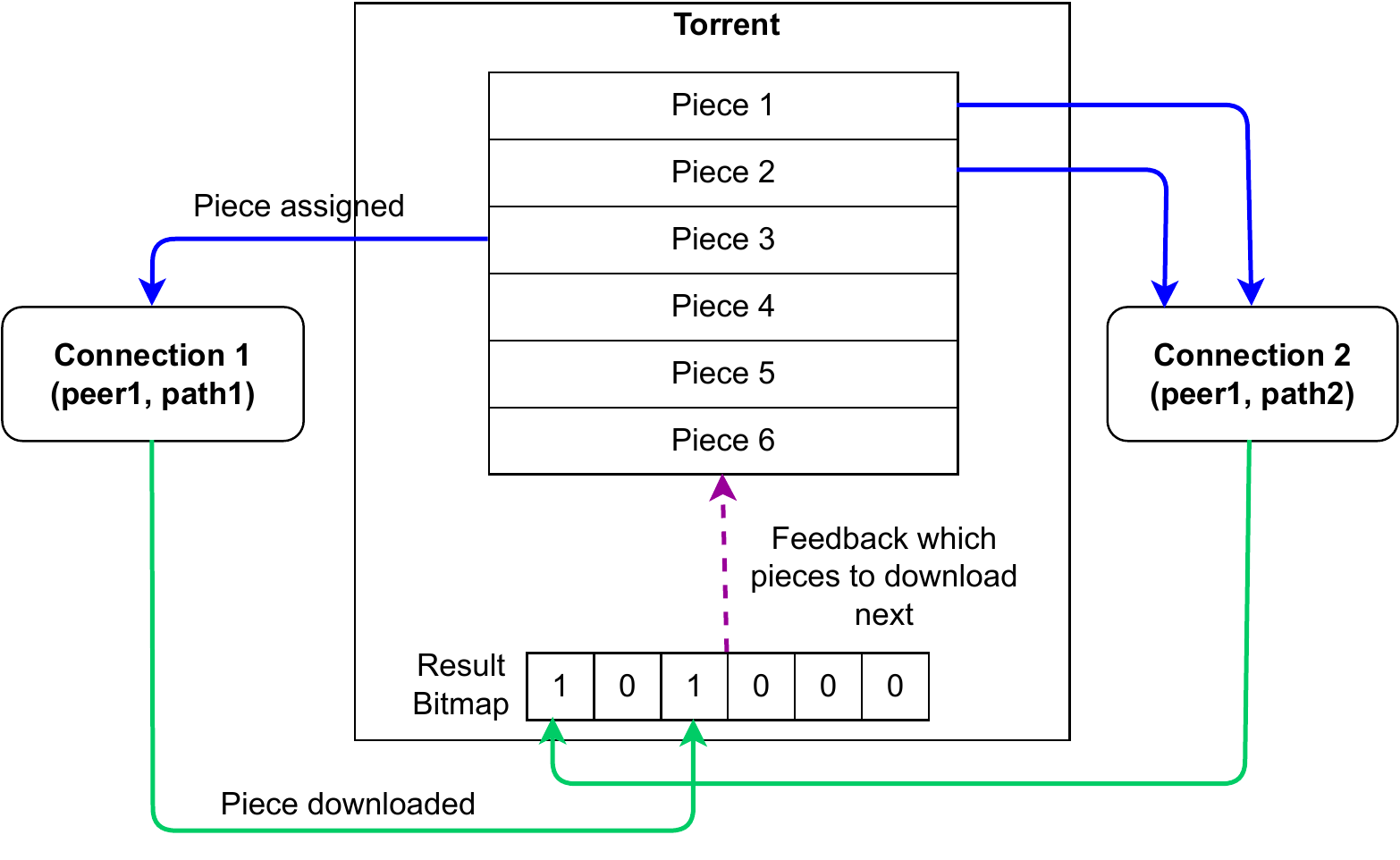}
		\caption{
			Multipath implementation by downloading a torrent file over multiple paths (connections).
		}
		\label{fig:torrentdownloadimpl}
	\end{figure}
\end{center}

We introduce this notion of path-level peers to a path-aware BitTorrent implementation which we will henceforth refer to as \emph{BitTorrent over SCION}, and which treats different paths to the same peer as several distinct, path-level peers.

Generally, peers that are returned from a tracker or that are added via static bootstrapping\footnote{I.e., providing a list of peer addresses as arguments to the client binary.} to BitTorrent are address-level peers, since they may not be SCION peers to begin with and may not have path information associated with them. Thus, to generate path-level peers from a given SCION address, an additional path lookup is required. For a resulting number of possible paths that are available to a peer, the same number of corresponding path-level peers can be generated.

BitTorrent over SCION is configured with an upper bound for the number of different path-level peers, as the default BitTorrent client is, too. Path-level peers are obtained from address-level peers through a path selection algorithm. 

After this obtaining of path-level peers, the usual BitTorrent P2P algorithms operate over each path independently. For each path-level peer, a QUIC\footnote{We choose QUIC, because TCP is currently not yet implemented for SCION.}\cite{langley2017quic} connection is established to ensure reliable transfer of data. Figure \ref{fig:torrentdownloadimpl} shows the piece download of a particular file. Each successfully established connection fetches piece information from a queue and requests the particular piece by sending request messages and wait for peers sending back the requested pieces. After each received piece, its integrity is verified. For this, a hash is computed and verified against the one referenced in the torrent file for that piece. Retrieved pieces are stored in main memory until the file is downloaded completely.

Since requests and retrievals of pieces over different connections are handled in their own dedicated threads, pieces can easily be downloaded concurrently, which speeds up the process and improves the overall download bandwidth. It is the responsibility of the main thread to iteratively check the result bitmap for missing pieces. Once all pieces are retrieved, the main thread closes all connections to all still connected peers that serve pieces of the current torrent and assembles the complete file to disk.

\subsection{Upload-based Disjoint Path Selection}

Although BitTorrent over SCION may simply use all available paths to each connected peer, there are reasons to limit the number of path-level peers. When this limit is set suitably high, all available paths to each peer are considered. However, multiple paths may share the same bottleneck, making it pointless to aggregate them in hope of increasing overall performance. Since each BitTorrent peer has an upper limit of outgoing connections to neighbours, a performance increase could also be expected by simply increasing this upper limit to exchange pieces with more peers. Consequently, using the shortest path or simply aggregating all paths does not promise to have significant impacts on BitTorrent over SCION's performance compared to path-unaware BitTorrent implementations. To address this, we use built-in SCION capabilities to implement an improved, \emph{disjoint} path selection strategy, aiming to avoid such shared bottlenecks. 

In BitTorrent over SCION, our improved path selection strategy relies on two core assumptions:
1) Different paths that share the same hop may also share a bottleneck at this hop and 2) Peers that offer pieces to others are in a good position for path selection decisions with knowledge about all downloading peers, allowing them to strategically distribute their outgoing traffic via disjoint paths. 

We implement a disjoint path selection strategy by searching for overlaps in all paths and discard all but one of the paths that share the same hops, following assumption 1). With respect to assumption 2), we decide to delegate path selection exclusively to the uploading peer. Both in combination promise to outperform na\"ive path selection approaches.

\begin{figure}	\includegraphics[width=0.5\textwidth]{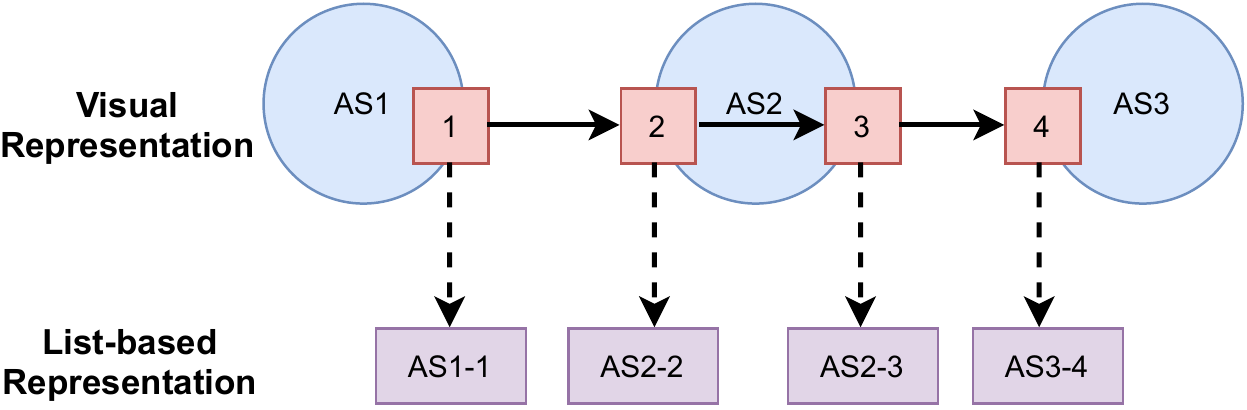}
		\caption{
			List-based representation of hop interfaces using unique ids to perform conflict detection.
	}
	\label{fig:aslist}
\end{figure}

As presented in Section \ref{sec:background}, a SCION path consists of multiple hops. Each hop contains ingress and egress interfaces of the respective AS. In SCION, the interfaces are represented as numeric IDs, that are unique within the given AS. The combination of the AS identifier with the interface number results in a globally unique \emph{interface ID}. In Figure \ref{fig:aslist}, we show an intuitive mapping of a visual representation of hops and their interfaces to a list of such interface IDs. The interface IDs are used to determine disjointness between paths by counting the number of identical interface IDs.

%

\begin{algorithm}
\caption{Disjoint path selection}\label{alg:two}
\KwData{$peers, maxOutgoingConns$}
\KwResult{$pathLevelPeers \gets [\:]$\;}
$allPaths \gets [\:]$\;
\For{$p \in peers$}{
    $paths \gets lookupPaths(peer)$\;
    $allPaths \gets append(allPaths, paths)$\;
}

\For{$path1 \in allPaths$}{
    \For{$path2 \in allPaths$}{
        \If{$path1 != path2$} {
            $confs \gets numConflicts(path1, path2)$\;
            $path.conflicts += confs$\;
        }
    }
}

$allPaths \gets sortByConflictsAndHops(allPaths)\;$

$i \gets 0$\;
\While{$i \le maxOutgoingConns$}{
    $pathLevelPeer \gets fromPath(allPaths[i])$
    $pathLevelPeers \gets append(pathLevelPeers, pathLevelPeer)$\;
    $i \gets i + 1$
}

\label{alg:disjoint-pathselection}
\end{algorithm}

Based on the interface IDs in the SCION paths, the uploading peer is able to perform disjoint path selection to all connected peers. To achieve this, an interested peer connects over the first available path to the uploading peer and waits for it to connect back. The uploading peer now applies its disjoint path selection and connects back to the interested peer over the selected path set. Algorithm \ref{alg:disjoint-pathselection} depicts the procedure of finding the least disjoint paths to all connected address-level peers and returning a proper list of path-level peers. At first, the paths to each address-level peers are determined in a loop and aggregated in the \emph{allPaths} variable. Afterwards, each path in allPaths is checked against all other paths calculating the number of conflicts (i.e. conflicting interface IDs), which is saved in the path. Next, the allPath list is sorted in ascending order by the number of conflicts and the number of hops. Finally, until the number of \emph{maxOutgoingConns} is reached, the algorithm iterates over allPaths and transforms each path into a path-level peer, which is stored in the return variable \emph{pathLevelPeers}.


\section{Benchmark Setup: A Representative Small-Scale Internet Testbed}
\label{sec:benchmark-setup}
\begin{center}
	\begin{figure*}
		\includegraphics[width=1.0\textwidth]{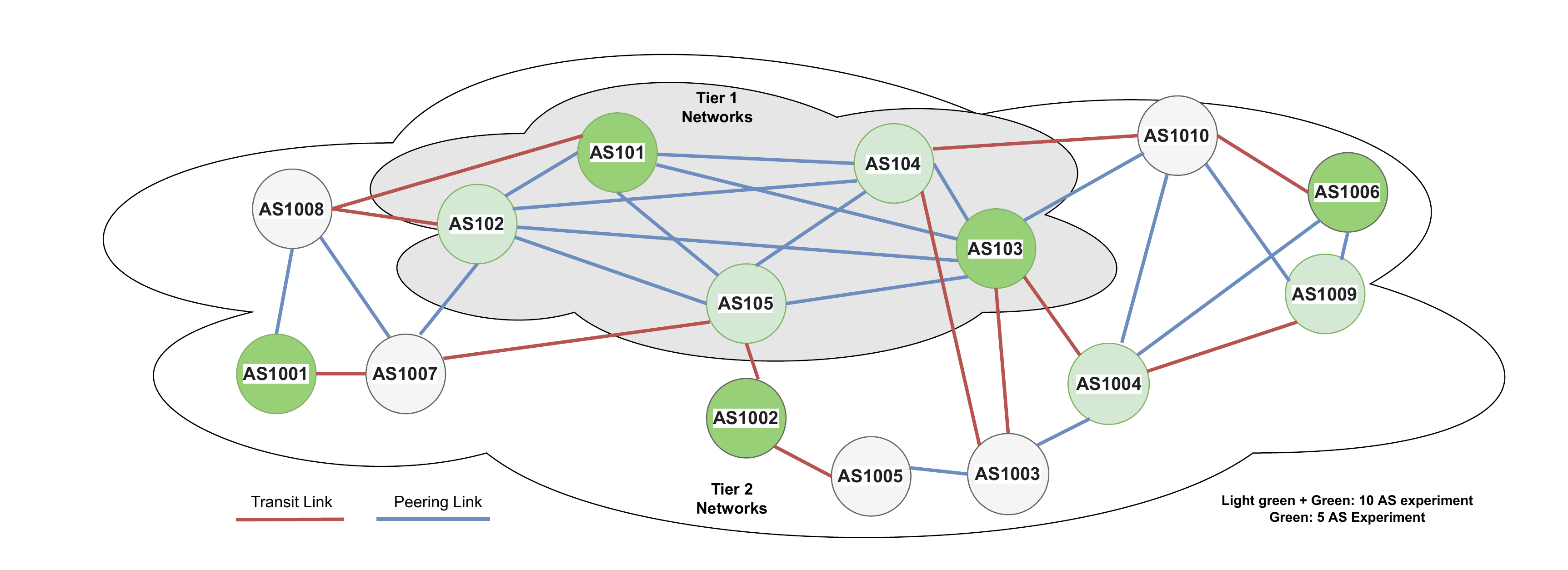}
		\caption{
			Virtualized testbed of a representative Internet topology
		}
		\label{fig:testbed}
	\end{figure*}
\end{center}

To compare our BitTorrent over SCION implementation against a BGP-based setup, we choose to run a virtualized network of multiple ASes that reflects the topology of the current Internet in a small-scale setup. Figure \ref{fig:testbed} shows the designed topology that we used to evaluate BitTorrent over SCION.

The topology consists of two core layers which represent Tier-1 and Tier-2 ASes in the Internet, connected via peering and transit links. With the growing popularity of IXPs, the number of Tier-3 ASes decreased significantly over time. Consequently, our testbed does not contain any Tier-3 ASes. The topology design follows actual AS and peering data, obtained from CAIDA data \cite{caidageo, caidarank, caidarel} and peeringdb \cite{peeringdb}, with randomized AS numbers. We limit the network link capacities to 15 Mbit/s for Tier-1 links and 10 Mbit/s for all remaining links to make all candidates network bound. Otherwise, the CPU could limit candidates resulting in potentially biased results. We derive two torrent networks from our proposed topology: The first network 5ASes consists of the 5 ASes marked dark green (AS102, AS1002, AS1004, AS103, AS1006) and the second network consists of the 10 ASes marked dark and light green (AS102, AS1002, AS1004, AS103, AS1006, AS1001, AS101, AS04, AS105, AS1009). ASes are connected either with transit or peering links. Our topology follows the \emph{valley-free} routing \cite{qiu2007toward}. Consequently, peering links can only be used by the peering neighbours and their customers. In BGP and BGP-M, this is implemented with prefix lists on each AS that has peering links. Since filtering support of peering links in SCION is currently in progress, we apply a static path filter to each AS to filter out valley-free violations. Since ISD's are a concept exclusively for SCION, we locate all SCION ASes in the same ISD, to achieve better comparability.

In our evaluations, we run three different BitTorrent candidates: The first one is BitTorrent over BGP, which is a usual BitTorrent client implemented in Go based on inter-domain routing performed by BGP. The second candidate is BitTorrent over BGP-M, which uses the same BitTorrent client but with configured load splitting in BGP in each AS. Despite other interesting approaches for load splitting, we apply ECMP-based load splitting for BGP-M in our testbed, since it is the most deployed approach in the current Internet. Our third candidate is BitTorrent over SCION, which implements the disjoint path selection based on the presented idea of path-level peers. 

The complete topology is running on a bare-metal server in a virtualized environment based on Docker. Each AS runs one or more containers (routers, hosts). Multiple ASes are connected via Docker Bridge Networks \cite{dockerbridge}. The server is running an Intel(R) Xeon(R) Gold 6150 CPU @ 2.70GHz with 36 threads and 500GB of main memory.

\section{Evaluation}
\label{sec:eval-large}
After presenting our virtualized Internet topology, we evaluate our BitTorrent over SCION approach in this section. 

\subsection{Terminology}
The following parameters are used to create and evaluate different BitTorrent experiments:
\begin{itemize}
    \item \emph{MaxPeers}: Number of available peers in the torrent
    \item \emph{OutgoingConns}: Number of outgoing connections (to peers) that each peer instantiates
    \item \emph{NumASes}: Number of involved ASes in the experiment
\end{itemize}

To design our experiments, we stick to findings from Hamra et al. \cite{hamra2007understanding} measuring BitTorrent's performance in real-world torrents. In most cases, MaxPeers is significantly higher than OutgoingConns, meaning each peer exchanges pieces with a subset of all availables peers. Hamra et al. show that setting OutgoingConns to the half of MaxPeers is a good tradeoff. Furthermore, the MaxPeers parameter changes over time in real-world torrents. This number increases often at the beginning of the torrent when many peers are interested in downloading the file and decreases after the majority of peers finished downloading and leave the torrent. We adapt this behaviour for our experiments.

In the following, we present the results of our two experiments: At first, we evaluate how heavy multipathing implemented in BitTorrent over SCION can aggregate bandwidth in the network that is not available for BitTorrent over BGP/BGP-M. Afterwards, we compare BitTorrent over SCION against BitTorrent over BGP/BGP-M with a varying number of OutgoingConns to evaluate the effect of our disjoint path selection approach.

\subsection{Bandwidth Aggregation}
In our first experiment, we run 20 BitTorrent peers in the torrent networks 5ASes and 10ASes exchanging a 100Mbyte file. By choosing 2 different torrent network sizes with a fixed-size topology, we can evaluate the impact of density of peers and the number of additional ASes that are not participating in the torrent, and consequently may provide additional capacities. We set the OutgoingConns parameter to infinity in this experiment (in detail it is set to the maximum number of path-level peers that one peer can connect to) to evaluate the maximum possible goodput that all candidates can achieve. 

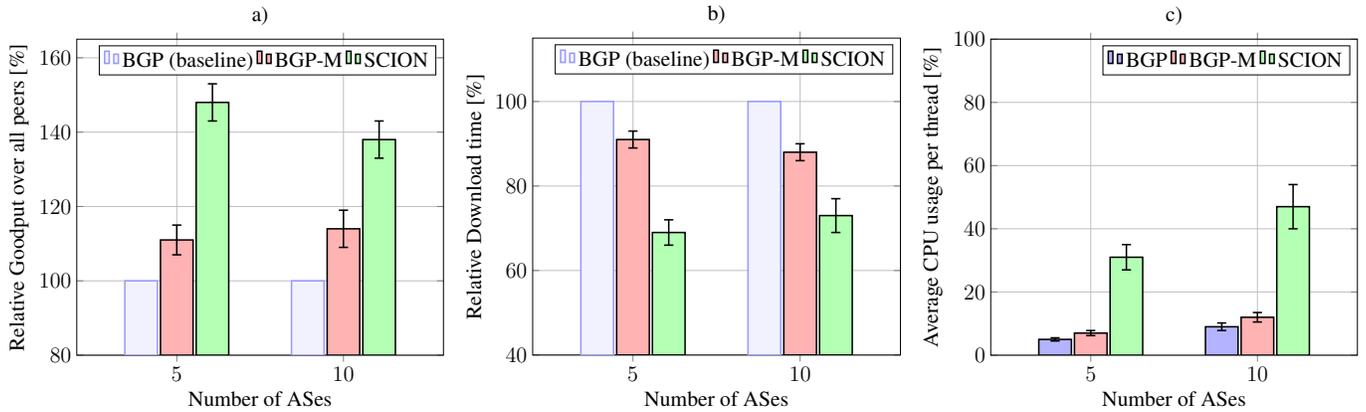
\begin{figure*}

\centering
\resizebox{0.33\textwidth}{!}{
\begin{tikzpicture}[font=\Large]
  \centering
  \begin{axis}[
        ybar,
        title={a)},
        bar width=0.75cm,
        ymin=80, ymax=165,
        legend style={
            legend columns=3
        },
        ylabel={Relative Goodput over all peers [\%]},
        xlabel={Number of ASes},
        xmin=2,
        xtick={5, 10},
        xmax=13,
    ]
    \addplot [fill=blue!5, draw=blue!40, error bars/.cd, y dir=both,y explicit, 
				error bar style={line width=1pt,solid},
				error mark options={line width=1pt,mark size=3pt,rotate=90}] coordinates {
      (5,100)
      (10,100) 
    };
   \addplot [fill=red!30, draw=black, error bars/.cd, y dir=both,y explicit, 
				error bar style={line width=1pt,solid,black},
				error mark options={line width=1pt,mark size=3pt,rotate=90}] coordinates {
     (5,111)+-(0,4)
     (10,114)+-(0,5)
 };
   \addplot [fill=green!30, draw=black, error bars/.cd, y dir=both,y explicit, 
				error bar style={line width=1pt,solid,black},
				error mark options={line width=1pt,mark size=3pt,rotate=90}] coordinates {
      (5,148)+-(0,5)
      (10,138)+-(0,5)
      };
    \legend{BGP (baseline),BGP-M,SCION}
  \end{axis}
  \end{tikzpicture}
}\hfill
\resizebox{0.33\textwidth}{!}{
\begin{tikzpicture}[font=\Large]
  \centering
  \begin{axis}[
        ybar,
        title={b)},
        bar width=0.75cm,
        ymin=40, ymax=115,
        legend style={
            legend columns=3
        },
        ylabel={Relative Download time [\%]},
        xlabel={Number of ASes},
        xmin=2,
        xtick={5,10},
        xmax=13,
    ]
    \addplot [fill=blue!5, draw=blue!40, error bars/.cd, y dir=both,y explicit, 
				error bar style={line width=1pt,solid},
				error mark options={line width=1pt,mark size=3pt,rotate=90}] coordinates {
      (5,100) 
      (10,100)
    };
   \addplot [fill=red!30, draw=black, error bars/.cd, y dir=both,y explicit, 
				error bar style={line width=1pt,solid,black},
				error mark options={line width=1pt,mark size=3pt,rotate=90}] coordinates {
     (5,91)+-(0,2)
     (10,88)+-(0,2)
 };
   \addplot [fill=green!30, draw=black, error bars/.cd, y dir=both,y explicit, 
				error bar style={line width=1pt,solid,black},
				error mark options={line width=1pt,mark size=3pt,rotate=90}] coordinates {
      (5,69)+-(0,3)
      (10,73)+-(0,4)
      };
    \legend{BGP (baseline),BGP-M,SCION}
  \end{axis}
  \end{tikzpicture}
}\hfill
\resizebox{0.33\textwidth}{!}{
\begin{tikzpicture}[font=\Large]
  \centering
  \begin{axis}[
        ybar,
        title={c)},
        bar width=0.75cm,
        ymin=0, ymax=100,
        legend style={
            legend columns=3
        },
        ylabel={Average CPU usage per thread [\%]},
        xlabel={Number of ASes},
        xmin=2,
        xmax=13,
        xtick={5,10},
    ]
    \addplot [fill=blue!30, error bars/.cd, y dir=both,y explicit, 
				error bar style={line width=1pt,solid},
				error mark options={line width=1pt,mark size=3pt,rotate=90}] coordinates {
      (5,5)+-(0,0.5)
      (10,9) +-(0,1.2)
    };
   \addplot [fill=red!30, error bars/.cd, y dir=both,y explicit, 
				error bar style={line width=1pt,solid},
				error mark options={line width=1pt,mark size=3pt,rotate=90}] coordinates {
     (5,7)+-(0,0.8)
     (10,12)+-(0,1.5)
 };
   \addplot [fill=green!30, error bars/.cd, y dir=both,y explicit, 
				error bar style={line width=1pt,solid},
				error mark options={line width=1pt,mark size=3pt,rotate=90}] coordinates {
      (5,31)+-(0,4)
      (10,47)+-(0,7)
      };
    \legend{BGP,BGP-M,SCION}
  \end{axis}
  \end{tikzpicture}
}

\caption{Comparison of BitTorrent over BGP/BGP-M and SCION in the 5AS and 10AS torrent network with 4 peers per AS. a) shows the aggregated goodput of all peers in \% with BGP as 100\% baseline, b) shows the download time in \% with BGP as 100\% and c) the aggregated CPU usage of all threads in \%}
\label{fig:bwaggr-results}

\end{figure*}

Figure \ref{fig:bwaggr-results}a) shows the aggregated goodput in percent of all peers for the three candidates, while BGP serves as baseline with 100\%. We decide to compare the goodput instead of the overall bandwidth, since SCION packets have a larger header than plain IP packets. 

For the torrent network 5ASes, we observe an aggregated goodput of 111\% for BGP-M compared to BGP, while SCION achieves 148\% compared to BGP. From these results, we observe that enabling multipath BGP via ECMP increases the overall goodput by 11\%. We conclude that in the 5ASes torrent network, the number of equal length BGP paths are comparatively low, leading to a small increase of goodput. However with BitTorrent over SCION's approach of path-level peers, a 48\% increase of goodput is achieved compared to BGP and a 33\% increase compared to BGP-M. In the 10ASes torrent network, we observe an increase of goodput for BitTorrent over SCION of still 38\% compared to BitTorrent over BGP, despite that the 10AS torrent network has a smaller number of not participating ASes that may provide additional network capacities. Consequently, BitTorrent over SCION is able to aggregate also heterogeneous paths and we confirm our hypothesis of BitTorrent over SCION's capability of aggregating bandwidth in the network that is unused in BGP/BGP-M.

To verify that increased goodput has a direct impact on the actual download time of peers, we measure the average download time of all peers in the 5ASes and 10ASes torrent networks, shown in Figure \ref{fig:bwaggr-results}b). Reflected by the lowest overall goodput, we observe the highest average download time for peers in BitTorrent over BGP, which is our baseline at 100\%. BitTorrent over BGP-M results in 91\% average download time for the 5ASes torrent network. In BitTorrent over SCION, the average download time is around 69\%. Also for the 10ASes torrent network, we observe that the goodput reflects the average download times, resulting in 73\% for BitTorrent over SCION and 88\% for BitTorrent over BGP-M. As expected, the increased goodput through aggregating unused capacities in the network directly translates into lower download times for peers, increasing the overall performance of the system. 

In addition to the goodput and download times, we also measure the overall CPU usage of all candidates running the 5ASes and 10ASes torrent network, shown in Figure \ref{fig:bwaggr-results}c). Since we observed an equal distribution of load over all threads, we present the CPU usage as average CPU usage per thread. While BitTorrent over BGP and BGP-M only use 5\% and 8\%, respectively, BitTorrent over SCION uses around 31\% of the available CPU usage for the 5ASes torrent network. We observe an expectable increase of resource usage of all candidates running 10ASes, with SCION using around 47\% of each thread. Despite the higher throughput, BitTorrent over SCION achieves, the increase of CPU usage is not inreasing in the same amount. We reason this increase by the open-source SCION stack \cite{sciongithub} (there also exists a closed-source SCION stack optimized for performance \cite{anapayaweb}). Especially the SCION Border Router implementation has potential to be optimized for performance, while the framework to route BGP and BGP-M (frrouter \cite{frrouter}) is heavily tuned. Consequently, we assume that using the closed-source SCION stack, we can decrease the CPU usage to a level comparable to BGP and BGP-M.

From this experiment, we confirm our hypothesis that BitTorrent over SCION can aggregate otherwise unused capacities in the network through multipath usage. We observe a significantly higher CPU usage for BitTorrent over SCION, which can potentially be strongly reduced by using the high-performance SCION stack, which is closed source.

\subsection{Peer Selection}
Always setting the OutgoingConns parameter sufficiently high may create unfair advantages for BitTorrent over SCION. Therefore, we compare all 3 candidates with a varying upper limit of OutgoingConns in this experiment. We again run 20 peers exchanging pieces of a 100Mbyte file in our 5ASes and 10ASes torrent network. As discussed before, setting OutgoingConns to the half of MaxPeers is a good tradeoff, we decide to vary OutgoingConns between 3 and 10. We expect that BitTorrent over BGP and BGP-M stagnates with a low number of OutgoingConns, meaning simply adding more connected peers to BitTorrent over BGP and BGP-M does not directly lead to improved performance, while BitTorrent over SCION handles an increasing number of OutgoingConns better.

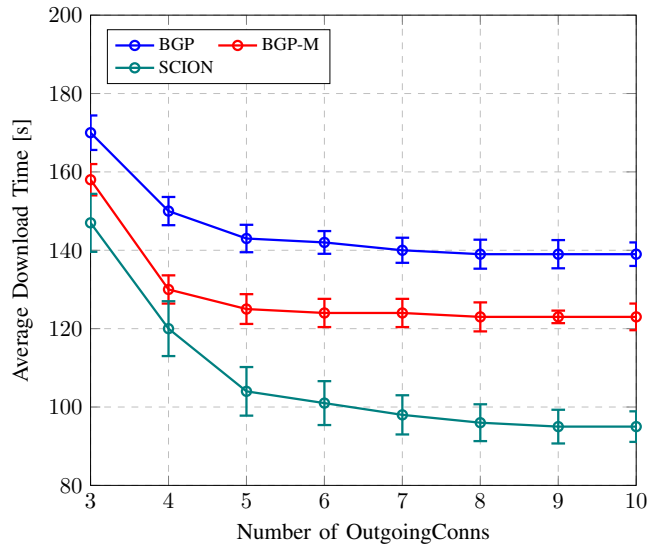
\begin{figure}
\resizebox{\columnwidth}{!}{
	\begin{tikzpicture}[font=\normalsize]
			\begin{axis}[
				xlabel={Number of OutgoingConns},
				ylabel={Average Download Time [s]},
				xmin=3, xmax=10,
				ymin=80, ymax=200,
				xtick={3,4,5,6,7,8,9,10},
				ytick={80, 100, 120, 140, 160, 180, 200},
				legend pos=north west,
				ymajorgrids=true,
				grid style=dashed,
				label style={font=\normalsize},
				tick label style={font=\normalsize},
				title style={font=\normalsize},
				legend style={nodes={scale=0.8, transform shape}},
				legend columns=2,
				]
				\addplot[
				color=blue,
				solid,
				mark=o,
				error bars/.cd, y dir=both,y explicit, 
				error bar style={line width=1pt,solid},
				error mark options={line width=1pt,mark size=3pt,rotate=90}
				]
				coordinates {
					(3,170)+-(0,4.4)(4,150)+-(0,3.6)(5,143)+-(0,3.5)(6,142)+-(0,2.9)(7,140)+-(0,3.2)(8,139)+-(0,3.7)(9,139)+-(0,3.6)(10,139)+-(0,3.0)
				};
				\addplot[
				color=red,
				solid,
				mark=o,
				error bars/.cd, y dir=both,y explicit, 
				error bar style={line width=1pt,solid},
				error mark options={line width=1pt,mark size=3pt,rotate=90}
				]
				coordinates {
					(3,158)+-(0,4.0)(4,130)+-(0,3.6)(5,125)+-(0,3.8)(6,124)+-(0,3.6)(7,124)+-(0,3.6)(8,123)+-(0,3.7)(9,123)+-(0,1.6)(10,123)+-(0,3.4)
				};
				\addplot[
				color=teal,
				solid,
				mark=o,
				error bars/.cd, y dir=both,y explicit, 
				error bar style={line width=1pt,solid},
				error mark options={line width=1pt,mark size=3pt,rotate=90}
				]
				coordinates {
					(3,147)+-(0,7.4)(4,120)+-(0,7.0)(5,104)+-(0,6.2)(6,101)+-(0,5.6)(7,98)+-(0,5.0)(8,96)+-(0,4.7)(9,95)+-(0,4.3)(10,95)+-(0,3.9)
				};
				
				\legend{BGP, BGP-M, SCION} 
			\end{axis}
		\end{tikzpicture}
}
\caption{
		Average download time in seconds for BitTorrent over BGP, BGP-M and SCION in the 5ASes torrent network.
	}
	\label{fig:disj-downloadtime-5}
 \end{figure}

Figure \ref{fig:disj-downloadtime-5} shows the average download time of all peers with a varying number of OutgoingConns for the 5ASes torrent network. We observe that BitTorrent over BGP and BGP-M start to stagnate after 5 OutgoingConns, while BitTorrent over SCION results in decreased download times until 8 OutgoingConns. With OutgoingConns greater than 7, the results are matching the ones presented in the bandwidth aggregation experiment. 

We conclude that  between 3 and 7 OutgoingConns, BitTorrent over SCION is still able to find disjoint paths between peers, while with more than 7 OutgoingConns, the additionally used path-level peers share bottlenecks. However, we assume that in real-world, Internet-scale torrent networks, the number of disjoint paths is significantly higher, leading to better results for higher numbers of OutgoingConns. 

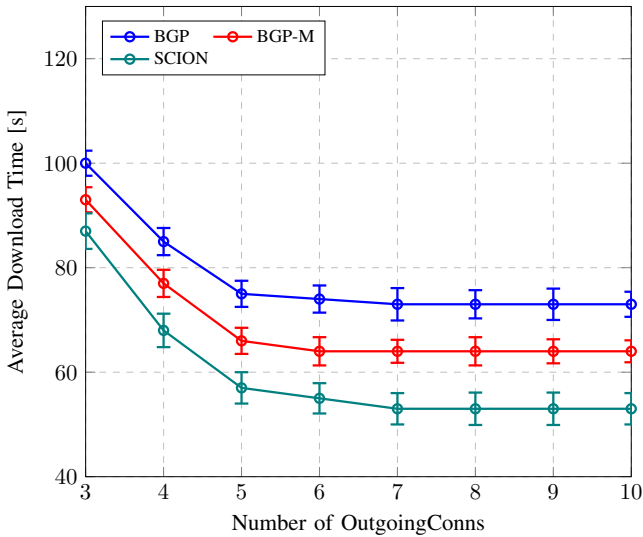
\begin{figure}
\resizebox{\columnwidth}{!}{
	\begin{tikzpicture}[font=\normalsize]
			\begin{axis}[
				xlabel={Number of OutgoingConns},
				ylabel={Average Download Time [s]},
				xmin=3, xmax=10,
				ymin=40, ymax=130,
				xtick={3,4,5,6,7,8,9,10},
				ytick={40, 60, 80, 100, 120},
				legend pos=north west,
				ymajorgrids=true,
				grid style=dashed,
				label style={font=\normalsize},
				tick label style={font=\normalsize},
				title style={font=\normalsize},
				legend style={nodes={scale=0.8, transform shape}},
				legend columns=2,
				]
				\addplot[
				color=blue,
				solid,
				mark=o,
				error bars/.cd, y dir=both,y explicit, 
				error bar style={line width=1pt,solid},
				error mark options={line width=1pt,mark size=3pt,rotate=90}
				]
				coordinates {
					(3,100)+-(0,2.4)(4,85)+-(0,2.6)(5,75)+-(0,2.5)(6,74)+-(0,2.6)(7,73)+-(0,3.1)(8,73)+-(0,2.7)(9,73)+-(0,3.0)(10,73)+-(0,2.4)
				};
				\addplot[
				color=red,
				solid,
				mark=o,
				error bars/.cd, y dir=both,y explicit, 
				error bar style={line width=1pt,solid},
				error mark options={line width=1pt,mark size=3pt,rotate=90}
				]
				coordinates {
					(3,93)+-(0,2.4)(4,77)+-(0,2.6)(5,66)+-(0,2.5)(6,64)+-(0,2.7)(7,64)+-(0,2.2)(8,64)+-(0,2.7)(9,64)+-(0,2.3)(10,64)+-(0,2.1)
				};
				\addplot[
				color=teal,
				solid,
				mark=o,
				error bars/.cd, y dir=both,y explicit, 
				error bar style={line width=1pt,solid},
				error mark options={line width=1pt,mark size=3pt,rotate=90}
				]
				coordinates {
					(3,87)+-(0,3.4)(4,68)+-(0,3.2)(5,57)+-(0,3.0)(6,55)+-(0,2.9)(7,53)+-(0,3.0)(8,53)+-(0,3.1)(9,53)+-(0,3.1)(10,53)+-(0,3.0)
				};
				
				\legend{BGP, BGP-M, SCION} 
			\end{axis}
		\end{tikzpicture}
}
\caption{
		Average download time in seconds for BitTorrent over BGP, BGP-M and SCION in the 10ASes torrent network.
	}
	\label{fig:disj-downloadtime-10}
 \end{figure}

In Figure \ref{fig:disj-downloadtime-10}, we show the average download time of all peers with a varying number of OutgoingConns for the 10ASes torrent network. We observe a high decrease of download times with less than 5 OutgoingConns for all candidates, while BitTorrent over SCION is still able to decrease the download time for up to 7 OutgoingConns. All three candidates have reached their minmum download times for greater than 7 OutgoingConns, in contrast to 5 OutgoingConns for the 5ASes torrent network. This is reasoned by the higher percentage of participating ASes in the torrent compared to the total number of ASes, and for BitTorrent over SCION especially by the lower number of additional network capacities.

From this experiment, we observe that BitTorrent over SCION also outperforms BitTorrent over BGP and BGP-M with a limited number of connected peers, disproving the intuitive argument that the path-level peer approach only works for a sufficiently high number of OutgoingConns. 

\color{black}
\section{Related Work}
\label{sec:relwork}
The mature P2P mechanisms behind BitTorrent have made the protocol an attractive target for networking research in the past. Ren et al. present \emph{TopBt} \cite{ren2010topbt}, an adaption of BitTorrent that uses proximities in addition to transmission rates to detect peers to collaborate with. Catro et al. propose \emph{BestPeer} \cite{castro2012bestpeer}, a peer selection algorithm that supports multipath in a multi-radio, multi-channel wireless mesh network.

Recent works cover analyses of BitTorrent's locality \cite{rumin2011deep, cuevas2013bittorrent, wang2009locality}, concluding that the majority of BitTorrent's traffic is still running globally. Furthermore, Decker et al. analyze behavioral patterns and topologies in existing torrent networks \cite{decker2013exploring}, while Cuevas et al. \cite{cuevas2013bittorrent} analyze how BitTorrent's locality impacts transit costs in existing networks.

A lot of research investigates IP multicast as an efficient way to distribute content to multiple peers without duplicating the traffic \cite{deering1989rfc1112}. Since IP multicast requires expensive dedicated support in network equipment, it has so far only seen localized deployment\cite{diot2000deployment, ratnasamy2006revisiting}. As an alternative to IP multicast, overlay approaches are considered: Bullet by Kosti{\'c} et al\cite{kostic2003bullet} is an overlay approach to efficiently distribute files from a single source to a large number of receivers. Also IPFS \cite{trautwein2022design} shares similarities with BitTorrent through its P2P based nature. Finally, Fujinoki et al. provide an approach to unlock private peering links for inter-domain routing in the Internet \cite{fujinoki2019beyond}, which provides interesting potential for our multipath approach.

Next to SCION, several other approaches to enable path control on the host exist. PathLet Routing by Godfrey et al. \cite{godfrey2009pathlet} is an approach based on segmentation of inter-domain routes into path fragments. Establishing multipath data transfer can also be realized completely on the application level: Yu et al. proposed mpath \cite{xu2012mpath}, an algorithm and implementation to leverage proxies to create multiple paths to a particular end host. 


To detect shared bottlenecks, different approaches have been proposed for MPTCP, some via active measurements \cite{wei2018shared, ferlin2016revisiting}, others via passive shared bottleneck detection \cite{hayes2014practical}. These approaches are constrained by an inherent lack of information about the actual path that the data uses through the Internet. With Espresso\cite{espresso}, Google presented a BGP-based approach for traffic distribution and bottleneck avoidance at the edge of their network, rather than on the endhosts. 

Finally, demonstrating SCION's high-performance capabilities, Neukom et al. propose Hercules \cite{neukom2020high}, a protocol for very high performance bulk data transfer over SCION and de Ruiter et al. present a SCION Border Router implementation in P4 \cite{de2021next}. 

\color{black}


\section{Conclusions and Future Work}
\label{sec:conclusion}

In this work, we developed the notion of \emph{path-level peers}, which allowed us to enhance BitTorrent with SCION support to add multipath features, with minimal modifications to the underlying file-sharing algorithms. Furthermore, we propose an algorithm for disjoint selection of path-level peers to improve the usage of network capacities. We evaluate BitTorrent over SCION in a virtualized inter-domain testbed comparing it to BitTorrent over BGP and BGP-M. We observe a 48\% improvement of goodput for BitTorrent over SCION compared to BGP and 38\% compared to BGP-M, which reflects in smaller average download times for participating peers. Furthermore, we show that our proposed disjoint path selection algorithm is able to improve traffic flow in the network with a low number of outgoing connections to unchoked peers. Consequently, we confirm our hypothesis that BitTorrent over SCION is capable of aggregating capacities in the network that are unused when BGP or BGP-M is utilized for inter-domain routing.

As future improvement for BitTorrent over SCION, we plan to extend the tracker implementation to pre-select particular path-level peers. Peers may actively communicate their selected path sets to the tracker, which can improve the location of shared bottlenecks based on the knowledge about path usage of all known peers. 

Furthermore, we plan to evaluate the impact of peers actively communicating the selected path set to other peers. We assume that peers can improve the location and avoidance of shared bottlenecks with this approach.

Finally, we plan to extend our disjoint path selection to allow multiple peers to reuse the same hops without creating shared bottlenecks, by observing variation in bandwidth to all connected peers when adding new paths containing shared hops. 

\bibliographystyle{splncs03}
\bibliography{bibliography}

\end{document}